\documentclass[11pt,a4paper,british]{article}
\usepackage[margin=3cm]{geometry}
\usepackage{authblk}
\usepackage{orcidlink}
\usepackage{rorlink}
\usepackage{microtype}
\usepackage[T1]{fontenc}
\usepackage{hyperref}
\usepackage{lmodern}
\usepackage[utf8]{inputenc}
\usepackage{parskip}

\hypersetup{
    colorlinks=true,
    linkcolor=teal,
    filecolor=magenta,      
    urlcolor=blue,
    pdftitle={Ensuring continued operation of INSPIRE as a cornerstone of
the HEP information infrastructure},
    pdfauthor={Sabine Crépé-Renaudin, Alexander Kohls, Micha Moskovic, Heath O'Connell, Kirsten Sachs, Jian Yu}
}
\begin{document}


\title{Ensuring continued operation of INSPIRE as a cornerstone of the HEP information infrastructure}

\author{
The INSPIRE Collaboration\\
Sabine Crépé-Renaudin\orcidlink{0000-0001-5980-5805}$^1$,
Alexander Kohls\orcidlink{0000-0002-3836-8885}$^2$,
Micha Moskovic\orcidlink{0000-0002-7638-5686}$^2$,
Heath B. O’Connell\orcidlink{0000-0002-1895-5310}$^3$,
Kirsten Sachs\orcidlink{0000-0002-6912-5800}$^4$,
Jian Yu\orcidlink{0000-0002-4434-7001}$^5$

\bigskip
This document has been endorsed by:\\
\bigskip
The INSPIRE Advisory Board\\
Alberto Accomazzi\orcidlink{0000-0002-4110-3511}$^6$,
John Beacom\orcidlink{0000-0002-0005-2631}$^7$,
Johannes Bluemlein\orcidlink{0000-0002-0565-4906}$^4$,
Brendan Casey\orcidlink{0000-0003-4260-3080}$^3$,
Kyle Cranmer\orcidlink{0000-0002-5769-7094}$^8$,
Yonit Hochberg\orcidlink{0000-0002-5045-2510}$^9$,
Kendall Mahn\orcidlink{0000-0002-0579-5928}$^{10}$,
Michelangelo Mangano\orcidlink{0000-0002-0886-3789}$^2$,
Michael Peskin\orcidlink{0000-0001-6403-6828}$^{11}$,
Jürgen Reuter\orcidlink{0000-0003-1866-0157}$^4$,
Qiang Zhao\orcidlink{0000-0001-8334-2580}$^5$,
Silvia Zorzetti\orcidlink{0000-0002-3208-3387}$^3$

\bigskip

Ursula Bassler\orcidlink{0000-0002-9041-3057}$^{12}$,
Florencia Canelli\orcidlink{0000-0001-6361-2117}$^{13}$,
Klaus Desch\orcidlink{0000-0001-5836-6118}$^{14}$,
Marek Karliner\orcidlink{0000-0001-7056-8898}$^{15}$\\
\bigskip
$^1$Université Grenoble Alpes, CNRS, Grenoble INP, LPSC-IN2P3, 38000 Grenoble, France \rorlink{https://ror.org/03f0apy98}\\
$^2$CERN, 1211 Geneva, Switzerland \rorlink{https://ror.org/01ggx4157}\\
$^3$Fermi National Accelerator Laboratory, Batavia, IL 60510, USA \rorlink{https://ror.org/020hgte69}\\
$^4$Deutsches Elektronen-Synchrotron DESY, Germany \rorlink{https://ror.org/01js2sh04}\\
$^5$Institute of High Energy Physics, CAS, Beijing, China \rorlink{https://ror.org/03v8tnc06}\\
$^6$Harvard-Smithsonian Center for Astrophysics, Cambridge, MA 02138, USA \rorlink{https://ror.org/03c3r2d17}\\
$^7$The Ohio State University, Columbus, OH 43210, USA \rorlink{https://ror.org/00rs6vg23}\\
$^8$University of Wisconsin-Madison, Madison, WI 53706, USA \rorlink{https://ror.org/01y2jtd41}\\
$^9$The Hebrew University, Jerusalem 91904, Israel \rorlink{https://ror.org/03qxff017}\\
$^{10}$Michigan State University, East Lansing, MI 48824, USA \rorlink{https://ror.org/05hs6h993}\\
$^{11}$SLAC National Accelerator Laboratory, Stanford, California 94309, USA \rorlink{https://ror.org/05gzmn429}\\
$^{12}$Laboratoire Leprince-Ringuet, F-91120 Palaiseau, France \rorlink{https://ror.org/058t6p923}\\
$^{13}$University of Zurich, 8057 Zürich, Switzerland \rorlink{https://ror.org/02crff812}\\
$^{14}$University of Bonn, 53113 Bonn, Germany \rorlink{https://ror.org/041nas322}\\
$^{15}$Tel Aviv University, Tel Aviv 69978, Israel \rorlink{https://ror.org/04mhzgx49}

}


\maketitle
The INSPIRE platform --- the most widely-used discovery service specifically tailored to the needs of researchers in High Energy Physics (HEP) --- has become a central component of the information infrastructure for the discipline. Despite this, INSPIRE’s continued sustainability is frequently endangered by resource constraints, recently made more acute by the loss of support from historical funders changing their research priorities. If the European particle physics community wishes to ensure INSPIRE’s long-term sustainability, the community should secure international support and ensure appropriate funding.\footnote{This document has been submitted to the European Strategy for Particle Physics Update 2026.}

\newpage

Particle physics is well known for its dependence on large-scale infrastructures like accelerators and detectors. Often overlooked, but equally crucial to enable the work of researchers, is the trustworthy infrastructure allowing them to access and navigate scientific information, and give visibility and credit to their scientific results.

INSPIRE has been a cornerstone of the information infrastructure for
HEP.\footnote{see the IUPAP Commission 11 "Support Letter on behalf of the High Energy Physics (HEP) Information Infrastructure", also submitted to the ESPPU, for a more general overview.}  First of all, it is the primary resource for HEP researchers
seeking accurate scholarly information.  Built up over decades,
carefully curated, and responsive to advice from its users, INSPIRE
contains an entry for every relevant paper in HEP and substantial
amounts of the literature in associated fields such as astrophysics
and statistics. Papers can be searched by keywords and through trains
of citation, offering a complete bibliography for any research topic.
Physicists around the globe make 25k visits to the service each day,
on average; 42\% of these visits originate in Europe. INSPIRE
integrates with the other prominent information source in HEP,
including arXiv, scientific publishers, the Particle Data Group,
HEPData, and ORCID.  INSPIRE also hosts the HEP community's
searchable lists of conferences and online seminars. Increasingly,
conference proceedings are not published but only maintained on web
sites. INSPIRE makes these resources accessible. 

Beyond this basic role, the community's view of INSPIRE as a complete
and trusted source gives it a role in career-related tasks for HEP
scientists.  It is used by younger scientists to compile CVs for job
applications, by senior scientists writing evaluation letters or
considering candidates for hiring, and by everyone in assembling
research grant applications.


Besides being a tool used on a daily basis by many HEP researchers, INSPIRE also stands out in the wider scholarly infrastructure landscape. Firstly, through its scope. Instead of trying to delineate its domain on an institutional or geographic basis (as institutional or national repositories do), or attempt to cover all of science (like generalist scientific databases) or a freely available subset (for Open Access repositories), INSPIRE is focused on a specific field which allows it to go deeper in its coverage (including for example theses and conference proceedings), be more accurate through more tailored automated processes and manual corrections, and make organisational decisions that are more aligned with the wishes of the community. To illustrate this, one example with far-reaching consequences is its unconventional handling of preprints: instead of considering them as completely separate papers from the journal publications as other platforms would do (or even outright ignore them), INSPIRE combines both into single records. This offers the benefit of having much more accurate citation counts (only a single citation is counted independently from the citing paper referring to the preprint, the published paper, or both) but on the converse requires more work to identify matching pairs and combine the sometimes conflicting information from the different sources.

Secondly, and in contrast to other platforms offering similar services, it is not operated by a for-profit entity charging hefty subscription fees to academic institutions (e.g., Web~of~Science, SCOPUS, dimensions) or as non-revenue generating aside to a successful company’s main business (Google~Scholar\footnote{It should be noted that Google Scholar gets some of its information from INSPIRE, so its contents will get worse if INSPIRE were to cease operations.}). Instead, INSPIRE is made available for free to the global HEP community via a collaboration of several scientific institutions with a deep understanding of the community needs and the success of HEP research at heart. Originally established as a collaboration between SLAC~(USA) and DESY~(Germany) in 1974 (and called SPIRES back then{\interfootnotelinepenalty10000 \footnote{For more details on the history of INSPIRE, see Hecker, B. Four decades of open science. \href{https://doi.org/10.1038/nphys4160}{Nature Phys 13, 523–525 (2017)}.}}), by the end of the 2010s, the INSPIRE collaboration had expanded also to CERN, Fermilab~(USA), CNRS~IN2P3~(France) and IHEP~(China). Its members provide in kind contributions for the required computing infrastructure (mostly at CERN) and staffing (from all collaboration members), comprising approximately 15 full-time equivalents (FTEs) across the whole collaboration.

This staffing level is the minimum of what is necessary to deliver a satisfactory service to the community, much of which is often not very visible. INSPIRE needs computer engineers to ensure the smooth operation and maintenance of the IT infrastructure and development of new functionality for both its user-facing website and the internal back office tooling, physicists to supervise the project, select the relevant content and categorize it, and, last but not least, curators (expert librarians) to ensure that the information in INSPIRE is complete and correct and offer support to users.

Historically, this was acknowledged by the management of the collaboration members, and INSPIRE has always received strong support and appreciation for its work. Unfortunately, over time this support has somewhat wavered. Some of the institutions in the collaboration increasingly steered their research programmes away from HEP and INSPIRE’s strength became a weakness, as they could no longer justify a significant investment into a field-specific platform for a field that itself wasn’t among their priorities. In 2021, SLAC ceased all of its activities in the collaboration and in 2024, DESY decided to drastically reduce its contribution. In the first case, the collaboration managed to cope by reducing some non-essential tasks and redistributing the rest among the still active members; but in the second case INSPIRE needed to seek external support to ensure continuity of its operations. Two German institutions, TIB --- Leibniz Information Centre for Science and Technology, and the Max Planck Digital Library (MPDL), stepped in to cover for DESY’s former duties for a limited duration of two years; STFC~(UK) and INFN~(Italy) have expressed interest in contributing on a more reduced scale,\footnote{A pilot programme is currently ongoing with STFC.} although these have yet to culminate into formalized commitments.

On a longer timescale, there is currently no guarantee that INSPIRE will manage to reach the staffing threshold allowing its operation  at a level that provides sufficient value to the HEP community. To reduce staffing needs the collaboration is investigating ways to increase automation, which seems promising thanks to the recent spectacular improvements in Machine Learning and in particular Large Language Models (LLMs); however, these require resources for the research and development stage as well as continuous quality management by experts. 

Despite this difficult situation, the INSPIRE team is committed to evolving the services it provides to answer to changing community practices. For instance, with the COVID pandemic and ensuing travel restrictions, online seminars that could be attended globally replaced local seminars for a geographically limited audience, prompting INSPIRE to allow users to list and browse seminars. More recently, INSPIRE has added a data collection to give visibility to research products beyond papers,\footnote{In a limited beta version for now, with HEPData as the only source of information. More information can be found in \href{https://blog.inspirehep.net/2025/03/introducing-the-inspire-data-collection-enhancing-open-science-through-dataset-discovery/}{the announcement on the INSPIRE blog}.} in line with the increasing importance of Open Science, which has also been recognized in the previous update of the European Strategy. 

For the future, INSPIRE has plans to expand this collection with additional sources for data and the inclusion of research software, and also add a collection of slides as an alternative to traditional conference proceedings. More radically, INSPIRE could leverage LLMs to make it much easier to find scientific products by offering natural language search capabilities complementing its traditional search syntax, thereby also offering a more user friendly interface for the next generation of physicists. Needless to say, those new tools, very valuable for the HEP community, can't be developed with the current resource perspectives.

In conclusion, INSPIRE plays a unique and major role for the global HEP research community that is expected to persist for the foreseeable future. It has exciting projects to develop its services that will benefit each researchers in the HEP community. As a consequence, the required funding to continuously sustain and develop the INSPIRE service should be secured through the commitment of the large laboratories, i.e., CERN and other major European HEP institutions, in partnership with other relevant institutions globally.\footnote{This need for funding has been recognized by the US HEP community. The \href{https://www.usparticlephysics.org/2023-p5-report/investing-in-the-future-of-science-and-technology.html\#67software-computing-and-data-science}{2023 P5 report} mentions "Area Recommendation 18: Increase targeted investments that ensure sustained support for key cyberinfrastructure components by \$8M per year in 2023 dollars. This includes widely-used software packages, simulation tools, information resources such as the Particle Data Group and INSPIRE, as well as the shared infrastructure for preservation, dissemination, and analysis of the unique data collected by various experiments and surveys in order to realize their full scientific impact."}

\vspace{10 mm}\textit{Acknowledgements:} Fermilab is managed by FermiForward Discovery Group, LLC under Contract No. 89243024CSC000002 with the U.S. Department of Energy, Office of Science, Office of High Energy Physics. 

\end{document}